\documentclass[fleqn,twoside]{article}
\usepackage{espcrc2}
\usepackage{amssymb}
\usepackage{graphicx}
\usepackage[figuresright]{rotating}

\newcommand{\AmS}{{\protect\the\textfont2
  A\kern-.1667em\lower.5ex\hbox{M}\kern-.125emS}}

\title{The ANTARES project }

\author{T. Montaruli\address[Bari]{Physics Department and I.N.F.N., 
        University of Bari, \\ 
        I-70126 Bari, Via Amendola 173, Italy},
        on behalf of the ANTARES Collaboration\address{
The ANTARES Collaboration list can be found at
http://antares.in2p3.fr
}}
\begin{document}

\begin{abstract}
\vspace{1pc}
The ANTARES deep-sea neutrino telescope will be located at a depth of 2400~m
in the Mediterranean Sea. Deployment of the detector will commence this 
Autumn and is expected to be completed by the end of 2004. With a surface 
area of the order of 0.1~km$^2$ it will be one of the largest European 
detectors. 
The aim of neutrino telescopes is to detect high-energy neutrinos
from astrophysical sources whilst also providing information on the 
early Universe. Successful operation of ANTARES in a deep sea environment
constitutes an important milestone towards the ultimate goal of
the construction of an underwater neutrino telescope at the cubic-kilometre 
scale.
The sky coverage of astrophysical sources offered by a Mediterranean 
neutrino telescope is complementary to any similar device at the South Pole. 
The current status of the project is discussed and the
expected performance of the detector is described in the context of the 
scientific programme of the project which comprises 
astrophysical studies, dark matter searches and neutrino oscillations.
\end{abstract}

\maketitle

\section{Introduction}

A promising challenge for exploring the Universe is the
detection of high-energy ($\gtrsim 1$~GeV) neutrinos. Such neutrinos could be 
produced by powerful cosmic accelerators, for example those in 
supernova remnants, active galactic nuclei, compact binaries,
such as micro-quasars, and those producing gamma-ray bursts.
Neutrino astronomy complements high-energy gamma astronomy since
the early Universe cannot be probed with high-energy photons due to
photon-matter and photon-photon interactions - 
gamma rays of a few hundred TeV from the Galactic 
Centre cannot survive their journey to the Earth.
The weakly interacting nature of neutrinos, 
coupled with the fact that they point to their source of origin 
without deviation makes them unique `probes' with which 
to investigate regions at distances larger than 50~Mpc. 

The production of neutrinos in cosmic accelerators requires hadronic 
mechanisms to be active in the source. Whilst most of the observed 
high energy emissions to date can be explained by electromagnetic processes,
recent observations of gamma-rays up to 5~TeV from SNR RXJ1713-39 
by the Cherenkov telescope CANGAROO \cite{CANGAROO} suggest 
that electromagnetic
mechanisms are inconsistent with the data but $\pi^0$ decay models 
can readily explain the measured energy spectrum. 
There is, however, some debate on this result \cite{Pohl}.

The technique employed by neutrino telescopes is dictated by the small 
neutrino cross section and the large background due to atmospheric muons. 
Natural Cherenkov radiators, such as water or ice, provide a large active 
volume at reasonable cost and the indirect detection of neutrinos through 
muons produced in charged current interactions increase the 'target' region 
by the muon range. Detectors are constructed at large depths where the atmospheric 
muon flux is significantly reduced compared to that at the surface. 
Upward-going muons produced by neutrinos having crossed the Earth, are 
recognised as the products of neutrino interactions 
in or close to the instrumented region. 

The Cherenkov light emitted by charged particles in deep water or ice 
is detected using an array of photomultiplier tubes (PMTs) which are housed, 
together with some associated electronic components, 
in a high pressure-resistant glass sphere
known as an optical module (OM).
The muon direction and energy are measured using the arrival times and 
amplitudes of the PMT pulses.
The detector sensitivity increases with energy due to the increase
in the $\nu$-N cross section, the longer muon range and
the increase in the amount of emitted Cherenkov light through secondary 
particles.

\section{Status of the ANTARES project}  
The ANTARES (Astronomy with a Neutrino Telescope and Abyss environmental 
RESearch) project \cite{Proposal} started in 1996 and 
involves physicists, sea science experts and engineers from France, Germany, 
Italy, Russia, Spain, The Netherlands and the United Kingdom. 
The ANTARES location (42$^{\circ}$ 50'N, 6$^{\circ}$ 10'E) 
coupled with the Earth's rotation gives an annual
sky coverage of about 3.5$\pi$ sr. The instantaneous overlap with the
AMANDA experiment, located at the South Pole \cite{AMANDA}, 
will be approximatively 0.5$\pi$ sr giving an integrated common coverage of
1.5$\pi$ sr. The Galactic Centre, 
an important potential source of high-energy neutrinos, 
will be visible for 67\% of the day.

The detector will be installed at 2400~m depth in the Mediterranean Sea,
37~km off-shore of La Seyne sur Mer, near Toulon (France). 
An extensive programme of site evaluation
has provided the relevant environmental parameters of the detector site.
Surveys have confirmed the suitability of the sea bed for detector 
deployment and sea currents have been measured, they average $\sim 6$~cm/s 
with a maximum observed value of 19~cm/s.
The average loss of light transmission through OM glass spheres due to 
bio-fouling and sedimentation has been determined by long-term 
measurements to be only $\sim 2\%$ at the equator of a glass sphere
one year after deployment. Moreover, this loss is a decreasing function 
of the zenith angle \cite{biofouling}. 
In order to be sure that the biofouling effect will be 
negligible for long-term measurements, 
ANTARES PMTs will be oriented downwards.

Water transparency affects the light detection 
efficiency and thus determines the natural
scale for spacing the OMs. 
The amount of light scattering predominantly 
affects track reconstruction and is a factor having a relevant 
effect on the angular resolution. 
For blue light the absorption length
is $\sim 55$~m while the effective scattering length (the
scattering length divided by $1-\langle\cos\theta\rangle$, where
$\theta$ is the scattering angle) is above the 100~m range.
At the ANTARES site the light scattering contribution is 
small compared to Antarctic ice. In fact,
95\% of the photons emitted by a pulsed isotropic LED source
at a distance of 24~m from a PMT are collected within 10~ns (30~ns) for blue 
(UV) light.  

\begin{figure}[htb]
\vspace{-11pt}
\includegraphics[width=20pc, height=16pc]{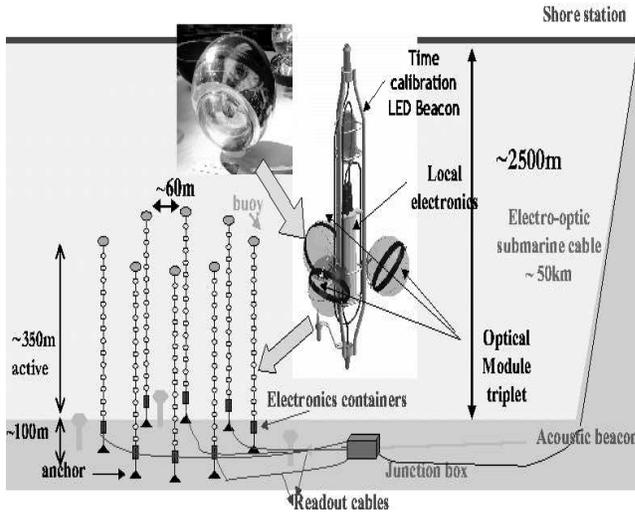}
\vskip -1 cm
\caption{Schematic view of the ANTARES detector showing an OM and some
details of a storey enlarged.}
\label{fig1}
\end{figure}

The detector, illustrated in Fig.~\ref{fig1}, consists of a 3-dimensional 
array of OMs \cite{OM} arranged in strings made of mechanically resistent
electro-optical cables. They will be anchored at the sea bed and
held taut by buoys at the top of the string.
At least 10 strings each equipped with 90 OMs arranged in 30 storeys 
are foreseen. Strings will be separated on the sea bed by 
approximately 60~m and storeys will be vertically separated
by 12~m.
Each storey is equipped with 3 OMs oriented at $45^{\circ}$ 
to the downward vertical. Readout electronics for each 
group of 3 OMs are located in a titanium container mounted just
above the OMs.
The 10'' Hamamatsu R7081-20 PMTs in the OMs are sensitive to 
single photons \cite{OM}. 
The front-end electronics of ANTARES OMs includes the Analogue Ring Sampler (ARS) 
ASIC \cite{Feinstein} which has been developed to process PMT signals and 
measures the arrival time and charge for single photoelectrons 
(99\% of the events) and the full pulse shape for larger photoelectron signals. 
The ARS has been shown to have an intrinsic time resolution 
of $\sigma = 0.3$~ns. The overall time resolution,
including the PMT transit time spread, is about $\sigma = 1.5$~ns. 
Signals are digitized, then transmitted via
optical fibres in cables.
The cables from individual strings will be connected (using a submarine) 
to a junction box at the end of an electro-optical cable which sends signals 
to a shore station where data are recorded. 
LED beacons (one per six storeys) will be able to emit precise
short pulses of blue light in order to calibrate the detector's timing.
An extra string, called the "Instrumentation String", 
will be devoted to the measurement of environmental parameters 
and will include, at its base, a laser beacon that will also contribute 
to the timing calibration.

Trigger logic and electronics, track reconstruction
and background rejection must take into account the presence of
background light. The background rate is seen to consist of two 
constant components: the rate due to 
the $\beta$-decay of $^{40}$K present in sea
water and a continuum rate that varies on a time scale of
several hours due to bioluminescence. 
The background light from these components 
produces an almost constant counting rate
of about 60~kHz on a 10'' PMT. 
Furthermore, short bursts of bioluminescence have been measured with
a rise time of a few milliseconds and a few seconds duration. 
These bursts induce a MHz level counting rate in a PMT which leads, 
on average, 
to less than 5\% dead time on any individual PMT with 
negligible correlation between storeys.

The ANTARES collaboration has successfully achieved the following:
\begin{itemize}
\item from Nov. 1999 to June 2000, a so-called "demonstrator string" 
(of a different mechanical design to the current design) was deployed 
at a depth of 1100~m and connected to shore with a 
37-km electro-optical cable.
This string was instrumented with 7 PMTs and more 
than 50000 7-fold coincidences were recorded,
zenith angles were reconstructed using times and amplitudes.
The shape of the zenith angular distribution of atmospheric 
muons was reproduced reasonably well, despite there being only
a small number of PMTs and a single string \cite{Kouchner}. 
After reconstruction, the data were seen to include
a $\sim 50\%$ contribution of multiple muons - as expected from such a 
shallow site. 
The demonstrator string allowed the ANTARES relative and absolute 
positioning to be tested with a system of acoustic rangemeters, compasses 
and tiltmeters. Relative distances between 2 elements 
were measured with 5~cm accuracy and absolute positioning was obtained with 
$\sim 1$~m accuracy. These values garantee a sufficiently accurate
positioning for astrophysics searches.
\item A test of OM sphere implosion was performed in June 2000 using a 
mechanical structure consisting of 2 mechanically complete but 
un-instrumented storeys.
One OM was artificially weakened and imploded at 
a depth of 2600~m. After the implosion, the other spheres and 
the electronics box in the same 
storey had also imploded, but the components in the adjacent storey and
the mechanical cable remained intact;
\item the electro-optical cable to carry electrical power and data
between the ANTARES site and shore
was successfully deployed in October 2001;
\item a mechanical test of a prototype string deployment
has been performed in November 2001; 
\item in April 2002 the industrial production of 900 Optical Modules started.
\end{itemize}
The planning of the experiment includes the following key dates:
\begin{itemize}
\item Autumn 2002: deployment of a prototype string containing 5 storeys 
(called sector string) and of the 
instrumentation string; 
\item December 2002: connection to the junction box
and start of data taking; 
\item June 2003: deployment of first two strings; 
\item December 2004: operation of the 10 string detector.
\end{itemize} 
\section{Expected performance}

The ANTARES scientific programme is mainly devoted to the detection of
neutrinos of astrophysical origin produced in point-like sources
or coming from the whole sky as a diffuse flux.
The neutrino signal can be detected above the background due to
atmospheric neutrinos at energies greater than 10 to 100~TeV as a result of
the harder neutrino spectrum expected from cosmic accelerators
compared to neutrinos from cosmic ray interactions in the atmosphere. 
Unfortunately, estimates of neutrino fluxes at large energies
are affected by an uncertainty of about 2 orders of magnitude on
the prompt neutrino contribution coming from charmed meson decays. 
Since their spectrum is harder than those of the pion/kaon component
of the atmospheric neutrino spectrum, prompt neutrino fluxes
could constitute a relevant source of background for the
cosmic neutrino diffuse flux search, while for point-like and transient sources
the knowledge of direction and time will help to suppress this background.

An intensive study to simulate the 10 string detector response to different
neutrino fluxes and to reconstruct events has been undertaken. The
muon production in the neutrino interaction has been simulated by using
deep inelastic scattering with CTEQ3-DIS parton functions \cite{Ghandi}. 
The Earth shadowing effect due to increasing neutrino cross-sections
is properly taken into account.
Muons are propagated to the detector using MUM
code \cite{MUM} taking account of different media 
(rock/water). 
The water properties are simulated including the scattering
of photons and a 60~kHz background count rate due to $^{40}$K decays. 
A simulation of the detector response including the behaviour 
of the PMT and front-end electronics is also performed.

The important parameters which characterise 
a neutrino telescope are its effective area (which includes reconstruction 
and selection efficiencies), 
the angular resolution and the energy resolution. The muon effective area,
a ratio of the selected event rate to the flux of 
incident muons, depends on the selection criteria used for any 
specific analysis. For point-like source searches, particularly when
a signal should be observed across the entire sky and not 
necessarily associated
to any known object, strict selection criteria are mandatory to keep 
only well reconstructed events. Thus, good pointing accuracy
and a good rejection of the background of atmospheric muons is necessary.  
Other searches, for instance for known transient sources, 
such as gamma-ray bursts, which are almost background free on timescales 
of hundreds of seconds, can be performed by replacing stringent 
cuts on the quality of the reconstruction by an angular cut around 
the presumed source.
In this case, the effective area calculation is made by requiring that the
angle between the direction of the reconstructed muon and the neutrino 
direction is less than a defined angular cut.

The effective area for the ANTARES detector when using these 
two approaches is illustrated below. 
Fig.~\ref{fig2} depicts the effective area of the 10-string 
detector for muon events 
as a function of neutrino energy 
assuming an isotropic distribution of neutrinos inducing muons.
%\footnote{ANTARES measures 'internal' events with 
%the interaction vertex inside the instrumented region and
%'external' events where the secondary is measured after propagating 
%through large distances. These topologies have been taken into account 
%to derive the detection efficiency and acceptance at all energies.
%For this reason the effective area is presented as a 
%function of the parent neutrino energy which is uniquely defined
%for both topologies.}. 
The solid line is obtained applying strict selections in order to have 
high-quality reconstructed events (see below for the corresponding angular 
resolution plot in Fig.~\ref{fig4}).
The upper dashed curve, which exceeds the geometrical area at high energies,
is obtained for reconstructed muon events within an angular error lower
than $1^{\circ}$ from the neutrino direction after relaxing quality cuts
(e.g. for transient source searches).
\begin{figure}[htb]
\vspace{-.5cm}
\includegraphics[width=18.5pc, height=15pc]{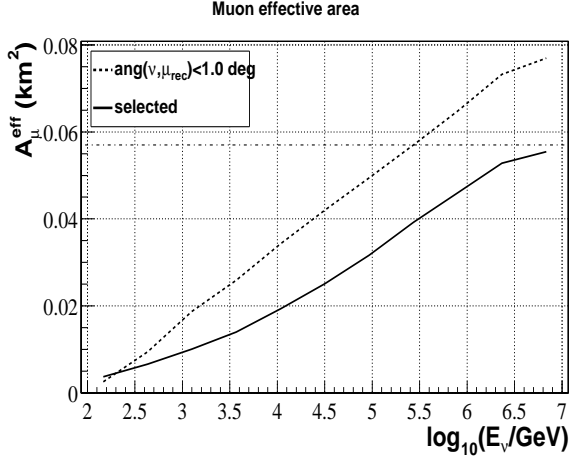}
\vspace{-1 cm}
\caption{Effective area versus neutrino energy for 
isotropically simulated events 
after strict reconstruction
quality cuts (solid line) and after relaxing the quality cuts but
requiring an angle between the reconstructed muon and the neutrino   
direction less than $1^{\circ}$ (dashed line). 
The geometrical instrumented surface is shown as a dashed-dotted line.}
\label{fig2}
\vspace{-.5cm}
\end{figure}

\begin{figure}[htb]
\vspace{-.5cm}
\includegraphics[width=20pc, height=15pc]{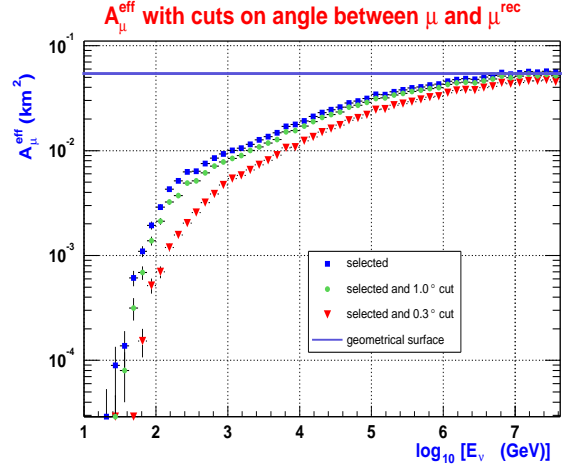}
\vspace{-1 cm}
\caption{Effective area vs neutrino energy after quality cuts:
squares are for all selected events, circles 
and triangles include the requirement
that the reconstruction error is less than $1^{\circ}$ and $0.3^{\circ}$, 
respectively.
The solid line represents the geometrical area.}
\vspace{-.5 cm}
\label{fig3}
\end{figure}

\begin{figure}[htb]
\vspace{-.5cm}
\includegraphics[width=18pc, height=14pc]{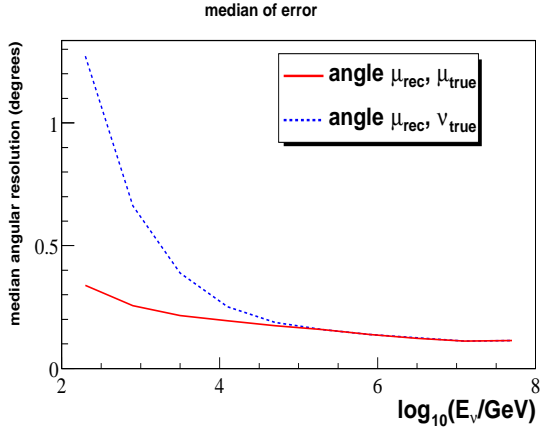}
\vspace{-1 cm}
\caption{Median angle of the distributions of the angle between
the reconstructed muon and the simulated muon (solid line)
or the simulated neutrino (dashed line)
versus neutrino energy.
Below 10 TeV the median angle between the muon and the neutrino is dominated 
by the kinematics of the interaction, while at larger energies it is limited
by the intrinsic angular resolution. The limiting value is about 
0.12$^{\circ}$.}
\vspace{-.5 cm}
\label{fig4}
\end{figure}

Fig.~\ref{fig3} illustrates how the effective area depends on the
intrinsic angular resolution for events selected by applying strict reconstruction
quality cuts. Together with the curve including all selected events, 
the curves for a reconstruction error of the muon (angle between 
the simulated muon and the reconstructed one) less than $1^{\circ}$ and 
than $0.3^{\circ}$
are shown demonstrating that most of the events are well reconstructed.
For a typical $E^{-2}$ cosmic accelerator neutrino spectrum \cite{GHS} 
96\% (72\%) of the events are reconstructed 
with an error less than $1^{\circ}$ ($0.3^{\circ}$). 

The intrinsic angular resolution of the telescope, defined as the median 
angular separation between the 'true' and the reconstructed muon track, 
has been estimated. 
In Fig.~\ref{fig4} the median value of the distribution of the angle 
between the reconstructed and simulated muon is shown as a solid line,
while the dashed line is the median angle between the reconstructed 
muon and the parent neutrino as a function of the neutrino energy.
The reconstruction quality cuts applied to obtain this 
plot are the same as those used for the solid line in Fig.~\ref{fig2}.
It should be noted that the improvement
of the angular resolution with increasing energy results in an increase of
the signal to noise ratio for neutrino astrophysics studies
at energies greater than 10~TeV. 
At these energies, the pointing accuracy is
only limited by the intrinsic angular resolution with a limiting value
of $\sim 0.12^{\circ}$. This estimate includes light scattering effects.
Good angular resolution facilitates powerful background rejection 
when searching for point-like sources. After an optimisation
of the signal ($E^{-2}$) to noise ratio the optimal search
cone is seen to have a half-width of $0.7^{\circ}$ around the known source
direction or around any measured neutrino event when looking for
clusters around it. Furthermore, it is found that the 'optimal' grid 
in the sky should have $70 \times 70$ bins in right ascension and
declination.\footnote{While the background is expected to 
be constant in right ascension, it depends on declination; hence the binning
in declination is variable in order to have constant background.} These
results can be compared to those from AMANDA B-10 which report an optimal
angular binning of $12^{\circ} \times 12^{\circ}$ \cite{Cowen}. 

\subsection{Sensitivity to point-like sources and to diffuse fluxes} 
\begin{figure}[htb]
\vspace{-.5cm}
\includegraphics[width=20pc, height=13.5pc]{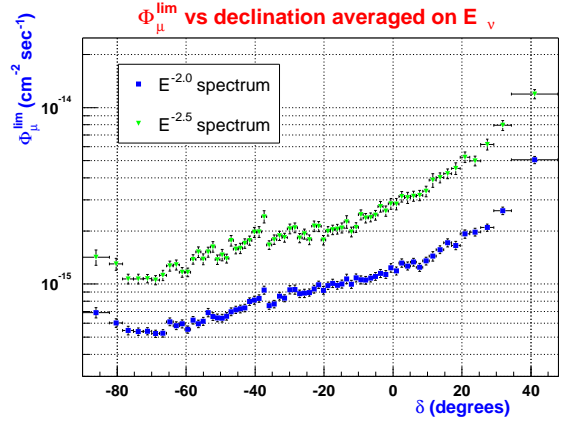}
\vspace{-1 cm}
\caption{Muon flux upper limits for two power law $\nu$ spectra 
(upper curve is for spectral index $\gamma = 2$   
and the lower curve is for $\gamma = 2.5$ of a $E^{-\gamma}$ spectrum) after
1 year of data taking. 
Neutrino energies are integrated between 10 and $10^7$~GeV.}
\vspace{-.5 cm}
\label{fig6}
\end{figure}

The effective area in Fig.~\ref{fig2} is directly correlated to the properties of the detector. To understand the response to a possible signal, 
the effective area has to be convoluted with a
model-specific neutrino flux. 
Taking as a reference an $E^{-2}$ signal spectrum, the response curve 
(differential rate of events versus energy) has a broad maximum
around neutrino energies of $10$ to $10^2$~TeV 
and around $3$ to $30$~TeV of the corresponding muon energies. 
The maximum of the response curve moves to lower energies for softer spectra.
The effective area averaged over an $E^{-2}$ spectrum is of the order of
22000~m$^2$ and decreases for softer spectra.
For point-like sources it is also important to consider
the dependence of the effective area on declination. 
%(ANTARES effective area 
%does not depend on right ascension).
This, together with an estimate of the background, allows 
the upper limits of the muon flux for any given type of source to be calculated. 
This study is performed considering physical backgrounds, including
atmospheric muons and neutrinos. 
The resulting upper limits in units cm$^{-2}$ s$^{-1}$ 
of muon flux for 1 year of data taking and for two neutrino models
are shown in Fig.~\ref{fig6} where the integration over neutrino energies
is made assuming $E_{min} = 10$~GeV.

To illustrate the expected sensitivity of ANTARES to point-like sources
the following example can be considered.
Recently a new calculation on neutrino fluxes from galactic micro-quasars
has been published \cite{mq} which predicts some promising expected fluxes. 
In particular, ANTARES expects to measure 6.5 and 4.3 events per year respectively for GX339-4 and SS433 
compared with a background of 0.3~events per year in a $1^{\circ}$ cone.

\begin{figure}[htb]
\vspace{-.5cm}
\includegraphics[width=18pc, height=13pc]{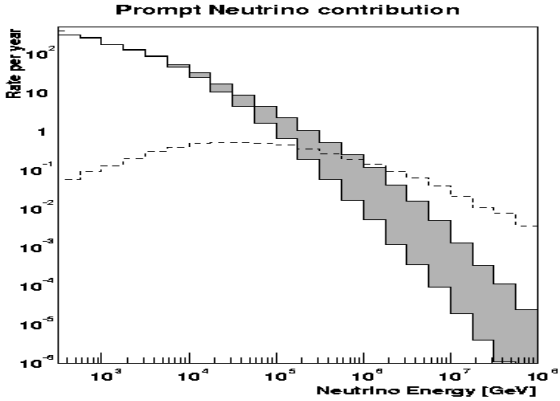}
\vspace{-1 cm}
\caption{Expected differential rate of events vs $E_\nu$ 
for atmospheric $\nu$s \protect\cite{Bartol}
and for the upper limit in \protect\cite{WB}. 
The shaded band represents the uncertainty on prompt $\nu$ models 
(references in \protect\cite{Costa}).}
\vspace{-.5 cm}
\label{fig7}
\end{figure}

Diffuse fluxes from astrophysical models are expected to exceed
the atmospheric neutrino background at energies above 10 to 100~TeV.
In Fig.~\ref{fig7} the expected differential event rate versus neutrino 
energy is shown for atmospheric neutrinos (Bartol flux \cite{Bartol})
and for the Waxman-Bahcall upper limit of $4.5 \times 10^{-8} E^{-2}$ GeV$^{-1}$ 
cm$^{-2}$ s$^{-1}$ sr$^{-1}$ 
\cite{WB} which is valid for extra-galactic sources transparent to 
nucleons. The shaded band at high energy
is derived from a set of prompt atmospheric neutrino models (references in 
\cite{Costa}). Currently the spread between the predictions covers 2 orders
of magnitude.

Fig.~\ref{fig7} indicates that the atmospheric neutrino background 
can be suppressed 
using the information on the reconstructed energy.
Below 100 GeV, the neutrino energy is calculated from the range of the muon
in the detector (for neutrino oscillation studies), while at energies 
above 100 GeV an energy estimator based on the PMT charge amplitudes 
currently allows muon energies to be reconstructed within a factor  
of 4 at 1~TeV decreasing to a factor of 3 at 10~TeV and 
reaching the value of 2 in the region 10 to $10^7$~TeV.
The ANTARES sensitivity to a diffuse neutrino
flux can thus be estimated by imposing a cut on the reconstructed energy which
minimizes the expected background with respect to the signal.
%This allows us to estimate the ANTARES 
%sensitivity to a diffuse neutrino flux 
%as an upper limit in the absence of any significant 
%excess of events above a given threshold. 
The predicted sensitivity is 
$E^2 d\Phi/dE = 10^{-7}$ GeV cm$^{-2}$ s$^{-1}$ sr$^{-1}$ above $10^{5}$ GeV,
one order of magnitude better than AMANDA B-10 \cite{Cowen} and Baikal \cite{Baikal} 
but still above the Waxman-Bahcall upper limit \cite{WB}.
\begin{figure}[htb]
\vspace{-.5cm}
\includegraphics[width=20.5pc, height=13.5pc]{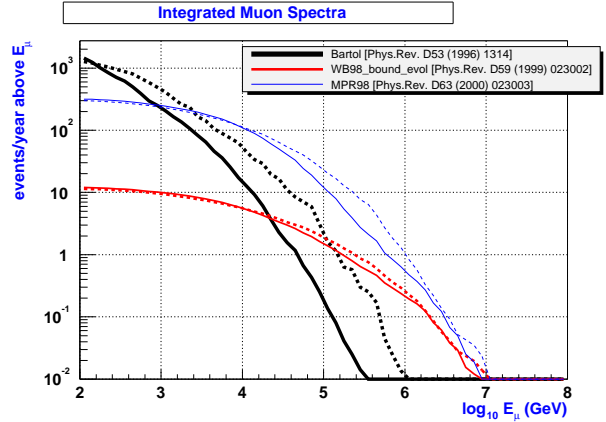}
\vspace{-1 cm}
\caption{Integrated rate of predicted (solid lines) and reconstructed (dashed lines) muons 
versus simulated and reconstructed 
muon energy respectively. The references for the models of atmospheric neutrino
fluxes \protect\cite{Bartol} and for upper limits to astrophysical neutrino sources 
are shown also in the plot (WB98~\protect\cite{WB} and MPR98~\protect\cite{MPR}).}
\label{fig8}
\vspace{-.5 cm}
\end{figure}
The expected event rates 
for the atmospheric neutrino background, 
for a flux equal to the Waxman-Bahcall \cite{WB} limit and 
for the more optimistic upper limit in reference~\cite{MPR}
are given in Tab.~1 for reconstructed muon energies greater than
10~TeV and 100~TeV.
The 'true' event rates for the same threshold on the true muon energy 
are also quoted as references.
The integrated event rates are illustrated in Fig.~\ref{fig8} 
where solid lines are the
expected spectra and dashed lines are the reconstructed ones.
The present ANTARES energy reconstruction reproduces harder spectra better 
than those of atmospheric neutrinos. For harder energy spectrum 
a small fraction of low energy events with over-estimated energy 
cannot contaminate significantly high energy events as it is the case for 
atmospheric neutrinos, which have the steepest energy spectrum.
This effect is enhanced by the fact that the energy resolution is a slightly
decreasing function of the energy.

\begin{table}
{\small
\begin{tabular}{|c|c|c|}
\hline
Flux & events/yr           & events/yr  \\
     &\small{$E_{\mu}^{reco}>10$ (100)TeV}&\small{ $E_{\mu}>10$ (100)TeV}\\ 
 \hline
Atm. & & \\
$\nu$s \protect\cite{Bartol}&44 (1.3)&12 (0.15)\\
Ref.~\protect\cite{WB}&5 (1.7)&5 (1.4)\\
Ref.~\protect\cite{MPR}&106 (20) &102 (11)\\
\hline
\end{tabular}}
\label{tab1}
\caption{Event rates above muon energy cuts at 10 and 100 TeV for
the Bartol atmospheric neutrino flux \protect\cite{Bartol}, for upper limits to extragalactic
neutrino fluxes in Ref.~\protect\cite{WB} and 
\protect\cite{MPR}. $E_{\mu}$ and $E_{\mu}^{reco}$ correspond to a cut
on the simulated energy and reconstructed energy respectively.}
\end{table}
\subsection{Sensitivity to WIMPS}
There is strong evidence in favour of a flat Universe 
with matter density $\Omega_{matter} 
\sim 0.35$. This result, combined with Big Bang nucleosynthesis 
(baryon density $\Omega_{baryon} \sim 0.05$) and with simulations of
structure formation, indicates that most of the dark matter in the Universe 
could be 
cold (non-relativistic at decoupling time) and non-baryonic.
Weakly Interacting Massive Particles (WIMPs) are considered among the most 
promising candidates, since a typical weak interaction cross-section would 
provide the relic density to close the Universe.
In a supersymmetric framework, assuming R-parity conservation, a 
theoretically-favoured candidate is the lightest supersymmetric particle,
which in many regions of the SUSY parameter space is the lightest neutralino
$\chi$. In minimal supergravity inspired models (mSUGRA), using
renormalization group equations and requiring radiative electroweak 
symmetry breaking, 5 parameters\footnote{These are: the common 
gaugino mass $m_0$, the common scalar mass
$m_{1/2}$, a common trilinear coupling A, the higgsino mass parameter sign 
sign($\mu$) and the ratio of the Higgs vacuum 
expectation values $\tan\beta$.}
determine the MSSM physics including the neutralino sector. 
%A theoretically-favoured candidate is the lightest supersymmetric particle, 
%which in many regions of the SUSY parameter space is the lightest neutralino
%$\chi$. 
%In the  minimal
%supergravity (mSUGRA) framework, five parameters determine the neutralino 
%sector.

Neutralinos in the galactic halo with typical velocities of few
hundreds of kilometres per second
could be slowed down sufficiently by scattering inside 
celestial bodies to become gravitationally trapped in their core.
Indirect dark matter searches aim to detect high-energy
neutrinos from WIMP annihilation with subsequent heavy quark and gauge boson 
decays in the core of the Sun, the Earth and
the Galactic Centre. 
Indirect detection is complementary to direct techniques which
observe the nuclear recoil from spin-independent WIMP-nucleon scattering in
scintillation detectors and in cryogenic phonon Ge and Si detectors. 
These two techniques have different backgrounds and systematic errors
and they have better sensitivity in different regions of parameter space.
For axially-coupled neutralinos indirect detection is
favoured because the Sun is largely made of hydrogen.
In addition, dark matter in the Galactic Centre is not accessible to
direct techniques. 

\begin{figure}[htb]
\vspace{-.5cm}
\includegraphics[width=20pc, height=18pc]{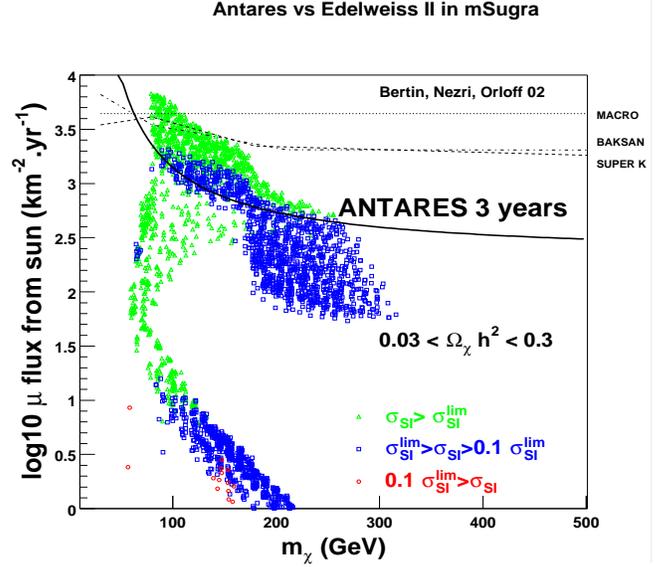}
\vspace{-1 cm}
\caption{
ANTARES (10 strings) 90\% c.l. sensitivity for neutralinos from the Sun 
for 3 years of data taking as a function of neutralino mass for
a hard neutrino spectrum ($\chi \chi \rightarrow WW$) compared to  
models described in \protect\cite{Lee,Nezri}. The minimum neutrino energy 
assumed is 5~GeV for the ANTARES sensitivity curve and for the models. 
$\sigma^{lim}_{SI}$ corresponds to the limit on the spin-independent
cross-section that can be set after about 3 years of
data-taking of Edelweiss II detector 
\protect\cite{Gerbier}. 
The upper limit is estimated using the Feldman and Cousins unified approach
\protect\cite{FC}. For comparison, limits from the indirect
experiments MACRO \protect\cite{MACRODM}, Super-Kamiokande 
\protect\cite{SK} and Baksan \protect\cite{Baksan} are superimposed.}
\label{fig9}
\vspace{-.5 cm}
\end{figure}

Indirect searches aim to detect a statistically significant 
excess against the background of atmospheric neutrinos pointing back towards
the Sun, the Galactic Centre and the nadir respectively.
Recent simulation results on ANTARES sensitivity 
have been presented in Ref.~\cite{Lee}.
In Fig.~\ref{fig9} the sensitivity curve (90\% c.l.) for 3 years of data taking
as a function of neutralino mass is shown. Models are described in 
\cite{Lee,Nezri}.
The search cone for neutrino events around the core of the Sun has been
optimized considering the angular distribution of the expected signal.
For comparison, limits from other experiments
are shown.

\subsection{Atmospheric neutrino oscillations}
\begin{figure}[htb]
\vspace{-.5cm}
\includegraphics[width=19pc, height=15pc]{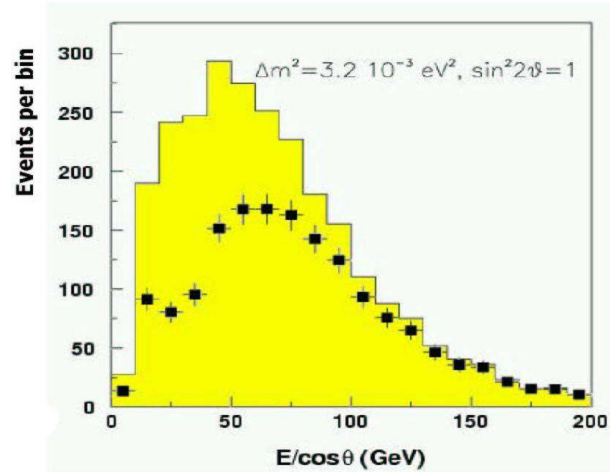}
\vspace{-1 cm}
\caption{Distribution of $E_{\mu}/\cos\theta$ 
which would be obtained by ANTARES 
after 4 years of data taking (data points) assuming
$\Delta m^2 = 3.2 \cdot 10^{-3}$ eV$^2$ and maximal mixing. 
Errors are statistical only.
The histogram is obtained assuming no oscillations.}
\label{fig10}
\vspace{-.5cm}
\end{figure}

Three different neutrino event topologies can be used to study 
neutrino oscillations: contained events with the vertex and the induced
muon inside the implemented region, semi-contained events with the 
vertex inside the detector but with the muon leaving it and
through-going events
with the vertex outside the detector.
Work is in progress to understand the systematic effects
and to improve the precision on the oscillation parameter measurement.
The measurement of the ratio, E/L, of the neutrino energy to the 
baseline length through the Earth for atmospheric neutrinos 
can provide evidence of neutrino oscillations.
The best estimator of this quantity is
$E_{\mu}/\cos\theta$, where $E_\mu$ is connected
to neutrino energy even though the hadronic component takes away part of the
energy in the interaction, and $\cos\theta_\mu$ is proportional to L,
even though at neutrino energies less than about 10~GeV the average 
angle between the muon and the neutrino is not negligible.
The visible muon energy is reconstructed from the muon range for contained
events selected by a containment cut based on PMTs along the track that 
have not been hit;
for partially-contained events an estimated correction is applied.
An analysis using the 720 single-string and 2100 multi-string contained and
semi-contained events expected in 4 yrs of data taking indicates 
that the oscillation pattern can be reconstructed in
the region allowed by Super-Kamiokande (see Fig.~\ref{fig10}). 

\section{Conclusions}
The ANTARES neutrino telescope is foreseen to be fully deployed by the end of 2004. 
The R\&D phase of the project, which started in 1996, has finished. 
During this phase a detailed assessment of the main requirements for an undersea neutrino telescope 
was made. The phase culminated in 1999 with the successful operation for several months of 
a demonstrator string which allowed the reconstruction of atmospheric muons. 
The project has now entered the construction and deployment phase for
a 0.1~km$^2$ scale detector. An electro-optical cable
has been successfully deployed and a prototype string and 
a string dedicated to environmental parameter measurements will be
deployed during this Autumn. 

Simulations tools are providing relevant information concerning the
scientific programme of the experiment, namely neutrino
astrophysics, dark matter and neutrino oscillations.
The predicted ANTARES sensitivity to fluxes from cosmic accelerators are
about an order of magnitude better than results presented to date by other
neutrino telescopes. 

%\vspace{-.5 cm}

\end{document}